\documentclass[epj,final]{svjour}
\usepackage[dvips]{epsfig}
\usepackage{graphicx}
\usepackage{amssymb}
\usepackage{dcolumn}
\usepackage{bm}

\newcommand{\LnCTO}{$Ln_{2/3}$Cu$_3$Ti$_4$O$_{12}$ }
\newcommand{\LnCRO}{$Ln$Cu$_3$Ru$_4$O$_{12}$ }
\newcommand{\LaCTO}{La$_{2/3}$Cu$_3$Ti$_4$O$_{12}$ }
\newcommand{\CeCTO}{Ce$_{1/2}$Cu$_3$Ti$_4$O$_{12}$ }
\newcommand{\PrCTO}{Pr$_{2/3}$Cu$_3$Ti$_4$O$_{12}$ }

\newcommand{\EuCTO}{Eu$_{2/3}$Cu$_3$Ti$_4$O$_{12}$ }
\newcommand{\GdCTO}{Gd$_{2/3}$Cu$_3$Ti$_4$O$_{12}$ }

\newcommand{\DyCTO}{Dy$_{2/3}$Cu$_3$Ti$_4$O$_{12}$ }
\newcommand{\HoCTO}{Ho$_{2/3}$Cu$_3$Ti$_4$O$_{12}$ }
\newcommand{\ErCTO}{Er$_{2/3}$Cu$_3$Ti$_4$O$_{12}$ }
\newcommand{\TmCTO}{Tm$_{2/3}$Cu$_3$Ti$_4$O$_{12}$ }
\newcommand{\YbCTO}{Yb$_{2/3}$Cu$_3$Ti$_4$O$_{12}$ }

\newcommand{\ACBO}{$AC_3B_4$O$_{12}$ }
\newcommand{\ACTO}{$A$Cu$_3$Ti$_4$O$_{12}$ }
\newcommand{\CCTO}{CaCu$_3$Ti$_4$O$_{12}$ }
\newcommand{\CCRO}{CaCu$_3$Ru$_4$O$_{12}$ }

\begin{document}

\title{On the magnetism of Ln$_{2/3}$Cu$_3$Ti$_4$O$_{12}$ (Ln = lanthanide)}

\author{A.~Dittl\inst{1}, S.~Krohns\inst{1}\fnmsep\thanks{\email{stephan.krohns@physik.uni-augsburg.de}}, J.~Sebald\inst{1}, F.~Schrettle\inst{1},
M.~Hemmida\inst{1}, H.-A.~Krug~von~Nidda\inst{1},
S.~Riegg\inst{1}, A.~Reller\inst{2}, S.~G.~Ebbinghaus\inst{3}, and
A.~Loidl\inst{1}}


\institute{Experimental Physics V, Center for Electronic
Correlations and Magnetism, University of Augsburg, 86135 Augsburg,
\linebreak[4]Germany \and Resource Strategy, University of Augsburg, 86135
Augsburg, Germany \and Solid State Chemistry, Martin-Luther
University Halle-Wittenberg, 06120 Halle, Germany}



%

\date{\today}

\abstract{The magnetic and thermodynamic properties of the complete \LnCTO
series were investigated. Here $Ln$ stands for the lanthanides La, Ce, Pr, Nd, Sm, Eu, Gd, Tb, Dy, Ho, Er, Tm, and Yb.
All the samples investigated
crystallize in the space group $Im\bar{3}$ with lattice
constants that follow the lanthanide contraction. The lattice
constant of the Ce compound reveals the presence of Ce$^{4+}$ leading to the composition Ce$_{1/2}$Cu$_3$Ti$_4$O$_{12}$. From
magnetic susceptibility and electron-spin resonance experiments 
it can be concluded that the copper ions always carry a spin $S=1/2$ 
and order antiferromagnetically close to 25\,K. The Curie-Weiss
temperatures can approximately be calculated assuming a
two-sublattice model corresponding to the copper and lanthanide ions, respectively. It seems that the magnetic moments of the heavy rare earths are weakly coupled
to the copper spins, while for the light lanthanides no such coupling was found. The $4f$ moments remain paramagnetic down to the lowest temperatures, with the exception
of the Tm compound, which indicates enhanced Van-Vleck magnetism
due to a non-magnetic singlet ground state of the crystal-field
split $4f$ manifold. From specific-heat measurements we
accurately determined the antiferromagnetic ordering temperature
and obtained information on the crystal-field states of the
rare-earth ions. The heat-capacity results also revealed the presence of a
small fraction of Ce$^{3+}$ in a magnetic $4f^1$ state.}

\authorrunning{A. Dittl \textit{et al}.}
\titlerunning{On the magnetism of Ln$_{2/3}$Cu$_3$Ti$_4$O$_{12}$ (Ln = lanthanide)}
\maketitle

\section{Introduction}

Perovskite derived oxides of \ACBO type constitute a broad new
class of compounds with fascinating properties. When three
fourths of the $A$-site ions of the parent perovskite $AB$O$_3$
are substituted by a Jahn-Teller active ion, like Cu$^{2+}$ or
Mn$^{3+}$, a collective rotation of the $B$O$_6$ octahedra around the
crystallographic $(111)$ axis gives rise to a square-planar coordination of these ions. Thus, the cubic lattice parameter doubles along
all three directions resulting in an eight times larger unit cell. These compounds are described by the
stoichiometry $AC_3B_4$O$_{12}$. Here the $C$-site cations possess a square-planar oxygen coordination analogous
to the Cu--O planes in the high-$T_{\rm c}$ superconducting cuprates. In the \ACBO structure a large variety of cations, irrespective of the nominal
charge state, can be substituted in the icosahedral environment of
the $A$ site, namely monovalent Na$^+$, divalent Ca$^{2+}$,
Sr$^{2+}$, Cd$^{2+}$, as well as trivalent Y$^{3+}$ and any
lanthanide element from La$^{3+}$ to Lu$^{3+}$ and -- as will be
shown later -- even tetravalent Ce$^{4+}$. As mentioned above, the $C$
positions are occupied by Jahn-Teller active ions
which strongly distort the perovskite structure by concomitant
orbital order. Finally, the $B$ site can be occupied by a large
variety of transition- and main-group metal ions, like e.g.
Mn$^{3+}$, Fe$^{3+}$, Ti$^{4+}$, and Ru$^{4+}$, to name the most
prominent examples. For a review on this class of new materials
see Ref.~\cite{Vasilev2007}. Recently \CCTO gained considerable attention
with respect to reported colossal values of the dielectric
constants \cite{Homes2001,Subramanian2000,Sinclair2002,Lunkenheimer2004,Lunkenheimer2010}.
CaCu$_3$Mn$_4$O$_{12}$ is a semiconducting ferromagnet with an
ordering temperature as high as 360\,K revealing large
magneto-resistance effects \cite{Zeng1999,Weht2002}. Heavy-fermion
behavior \cite{Kobayashi2004,Krimmel2008,Krimmel2009} has been
reported for \CCRO as well as for \LnCRO ($Ln =$ La, Pr, Nd) \cite{Buettgen2010} and finally a temperature induced valence transition accompanied by a metal-to-insulator transition has been found in
LaCu$_3$Fe$_4$O$_{12}$ \cite{Long2009}.

Here we focus on the magnetic properties of the insulating \LnCTO
compounds, where $Ln$ stands for any rare-earth element except Lu and the radioactive Pm. The synthesis of these oxides has first been reported by
Deschanvres \textit{et al.} \cite{Deschanvres1967} and Bochu
\textit{et al.} \cite{Bochu1979}. For charge neutrality one third of the $A$ sites remains vacant. Dielectric properties of this class of materials were published by
Subramanian \textit{et al.} \cite{Subramanian2000,Subramanian2002},
Liu \textit{et al.} \cite{Liu2005} and Sebald
\textit{et al.} \cite{Sebald2009}, discussing the colossal values of the
dielectric constant similar to CaCu$_3$Ti$_4$O$_{12}$. However, to the best of our knowledge no reports are available on magnetisation and specific heat of \LnCTO where rare-earth
magnetic moments are introduced in addition to the Cu spins.
Usually the Cu$^{2+}$ subsystem in insulating \ACTO behaves
simple: As documented for CaCu$_3$Ti$_4$O$_{12}$, the copper ions are in a
$3d^9$ electronic configuration with spin $S = 1/2$. They mainly interact via
the neighboring titanium and oxygen ions by super-exchange
interactions, thereby constituting a Curie-Weiss temperature which
ranges between -34\,K and -41\,K, depending on synthesis
conditions \cite{Krohns2009,Brize2009}. \CCTO undergoes antiferromagnetic
(AFM) order with a collinear spin arrangement at $T_{\rm N} =
25$\,K \cite{Kim2002}. The onset of magnetic ordering is associated
with a well-defined lambda-type anomaly in the specific
heat \cite{Koitzsch2002}. As Curie-Weiss temperature and ordering
temperature are of the same order of magnitude, the copper spins obviously are not frustrated.

In the rare-earth compounds
the coupling between copper and rare-earth ions is expected to be
weak due to the small overlap of the $4f$ shell with the neighboring ions. Therefore to first approximation, the ordering temperature should be
independent of the rare-earth spins. For the same reason the coupling 
between the rare-earth ions will be even weaker and
the corresponding super-exchange is negligible. However, when substituting Ti
ions by Ru, the system undergoes a metal-to-insulator transition,
where the localized Cu moments vanish and the compounds become
metallic with heavy-quasiparticle properties \cite{Ramirez2004}.
For the Ti rich side thermoelectric power was found to be strongly
dependent on the $Ln$ ions indicating an increasing hybridization
between Cu $3d$ and Ru $4d$ electrons with decreasing radius
of the $Ln$ ion \cite{Terasaki2010}. From this
point of view it is important to systematically investigate
the magnetic properties of the pure \LnCTO compounds which will be
one starting point for further experiments with solid solutions between
titanium and ruthenium compounds.


In this work the structural data of polycrystalline\linebreak[4]\LnCTO
compounds, with $Ln =$ La, Ce, Pr,  Nd, Sm, Eu, Gd, Tb, Dy, Ho,
Er, Tm, and Yb are presented.
We studied the magnetic bulk properties by SQUID measurements, while
electron-spin resonance (ESR) experiments were utilized to
probe the local environment of copper spins and rare-earth
moments. The specific-heat measurements provided a precise
measure of the magnetic ordering temperature and of the
crystal-field states of the rare-earth ions.


\section{Sample preparation and experimental details}

Polycrystalline samples of \LnCTO were prepared by solid-state
reaction. CuO, TiO$_2$, and the binary oxides, $Ln_2$O$_3$ -- with
the exception of Pr$_6$O$_{11}$, CeO$_2$, and Tb$_4$O$_7$ -- were
mixed in corresponding molar ratios and were well ground in an agate
mortar. Before weighting, all lanthanide oxides were dried at
$900^{\circ}$C for 12\,h to remove any water from the samples. In
the same way, Pr$_2$O$_3$ was reacted to non-hygroscopic
Pr$_6$O$_{11}$. An excess of 0.3\,g - 0.4\,g CuO was added as flux
material. After completion of the reaction, this excess was
removed by washing the samples with hydrochloride acid (10\%) and
afterwards with deionized water. Before calcination, the samples
were pressed into pellets. Calcination in aluminum-oxide crucibles
was performed in two steps, first at $1000^{\circ}$C for 48\,h in
air with subsequent regrinding followed by a second heating to
$1025^{\circ}$C again for 48\,h in air.

\begin{figure}
\resizebox{0.95\columnwidth}{!}{\includegraphics{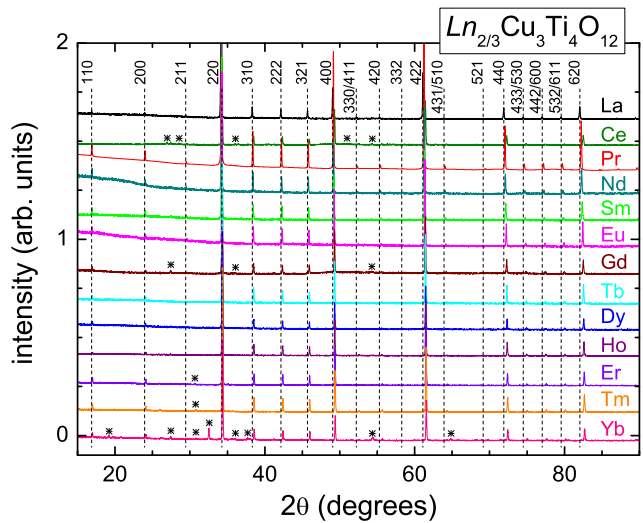}}
\caption{(Color online) Diffraction pattern of all
lanthanide compounds investigated in this work, vertically shifted for clarity. Vertical
dashed lines indicate the Bragg
reflections of \LaCTO and document the lanthanide
contraction. Asterisks denote impurity phases. The Miller indices
of all Bragg reflections are given on top of the figure.}
\label{Fig1}    
\end{figure}

The x-ray diffraction experiments were performed on a Seifert
3003 TT powder diffractometer using Cu-K$_{\alpha}$ radiation.
To provide qualitative information on the sample quality, Figure~\ref{Fig1}
shows the diffraction profiles of all lanthanide compounds investigated
from lanthanum (top) to ytterbium (bottom).
The Bragg reflections are indexed on top of the figure. Most of the \LnCTO
compounds are single-phase within the detection limit of XRD. Peaks, which are due
to impurity phases, are indicated by asterisks. Only \YbCTO and \CeCTO
exhibit a considerable amount of impurity phases and specifically
in the Yb sample a well developed Bragg reflection due to a
foreign phase shows up close to $33^{\circ}$. Traces of
impurity peaks with marginal intensity appear in the Gd, Er, and Tm
compounds. The dashed lines
indicate the scattering angles of the Bragg reflections of La$_{2/3}$Cu$_3$Ti$_4$O$_{12}$.
At large scattering angles the lanthanide contraction of the
series can easily be detected from the increasing shift of the Bragg reflections
to higher angles with increasing atomic number. Moreover, one observes that in\linebreak[4]\CeCTO
the Bragg peaks appear at too large scattering angles indicating a
smaller cell volume than expected. This is a hint, that Ce possibly is in a tetravalent
state, as already concluded in the early work by Bochu
\textit{et al.} \cite{Bochu1979}.


\begin{figure}
\resizebox{0.95\columnwidth}{!}{\includegraphics{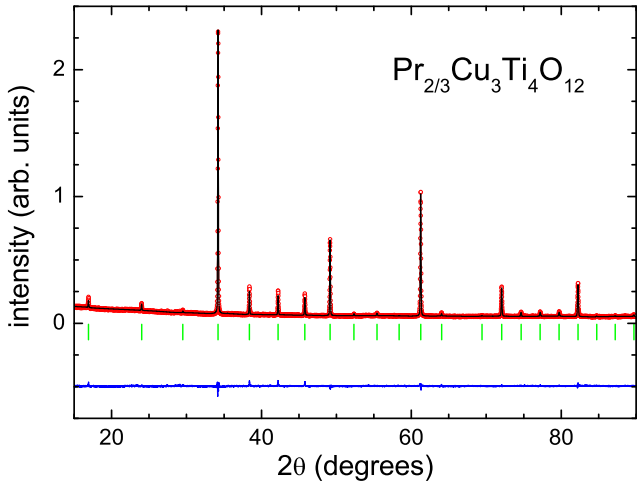}}
\caption{(Color online) Diffraction profile (empty
symbols) and Rietveld refinement (solid line) for
Pr$_{2/3}$Cu$_3$Ti$_4$O$_{12}$. The vertical bars indicate the
angular positions of the expected Bragg reflections. The difference
pattern between calculated and measured profile at the
bottom is shifted by -0.5 for clarity.}
\label{Fig2}
\end{figure}

To provide quantitative information on lattice constants and on
sample quality, all diffraction patterns were refined within space group $Im\bar{3}$ using the
Rietveld method, resulting in
$R$ values well below 5\%. As
a representative example, Fig.~\ref{Fig2} shows the diffraction
profile including the Rietveld refinement for Pr$_{2/3}$Cu$_3$Ti$_4$O$_{12}$.
The angular positions of the expected Bragg reflections are indicated
by vertical bars. The difference pattern proves the excellent fit and, thus, the phase purity of the sample.



\begin{figure}
\resizebox{0.95\columnwidth}{!}{\includegraphics{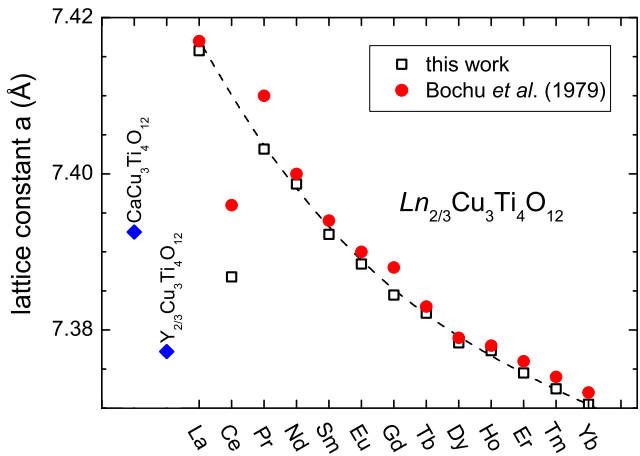}}
\caption{(Color online) Evolution of the lattice
constants of the lanthanide series (empty squares). The lattice constants of the
isostructural Ca and Y compound are also indicated. Expermimental results from Bochu \textit{et al.} \cite{Bochu1979} are plotted
for comparison (full red circles). The dashed line is drawn to
guide the eye and provides a smooth extrapolation of the trivalent
compounds.}
\label{Fig3}
\end{figure}

The evolution of the lattice parameter is documented in Fig.~\ref{Fig3}
for the complete lanthanide series in comparison with
published results by Bochu \textit{et al.} \cite{Bochu1979}. Except for \CeCTO and \PrCTO the agreement is excellent.
In addition the lattice constants of \CCTO and
Y$_{2/3}$Cu$_3$Ti$_4$O$_{12}$, which have been prepared in the
course of this work, are also shown. The cell parameters continuously decrease on increasing atomic number from 7.416\,\AA~
for La to 7.371\,\AA~for Yb due to the lanthanide
contraction. However, a decrease by 0.045\,\AA~is rather small, as the
ionic radii of the trivalent compounds (e.g., in octahedral symmetry)
decrease from 1.03\,\AA~for La$^{3+}$ to 0.87\,\AA~for Yb$^{3+}$. This
fact indicates that the available space for the $A$ cations is
rather fixed and determined from the strong tilting of the TiO$_6$
octahedra. 

The lattice constant for the Ce compound is too small
compared to the continuous evolution along the lanthanide series.
Often cerium is found to exhibit a $4f^0$ configuration and in
this case the ionic radius decreases from 1.01\,\AA~for the trivalent
ion to 0.87\,\AA~for Ce$^{4+}$. According to x-ray
diffraction the majority of the cerium ions seems to have
the valence $4+$ and the stoichiometry of this compound probably
has to be written as Ce$_{1/2}$Cu$_3$Ti$_4$O$_{12}$, if one
assumes strict charge neutrality. This means that in this
structure half of the $A$ sites are occupied by vacancies. So far
we were not able to synthesize \CeCTO in pure
form without impurity phases. On the other hand for the Ce compounds
investigated the impurity phase does not exceed 5\%.


In the further course of this work we investigated the magnetic properties and the specific heat. The dc magnetic susceptibility has
been studied in a temperature range $1.8 \leq T \leq 400$\,K utilizing a commercial Superconducting Quantum
Interference Device magnetometer (MPMS-5, Quantum Design). The heat capacity was measured in a
Physical Properties Measurements System (PPMS, Quantum Design) for
temperatures between 1.8\,K and 50\,K. In the specific-heat
measurements we focused on the low temperatures to get
information on the magnetic ordering, crystal-field excitations of
the rare-earth ions, and of the low-temperature phonon properties. The ESR experiments were performed in a Bruker ELEXSYS E500 CW
spectrometer at X-band frequency (9.4\,GHz). This ESR spectrometer
is equipped with a continuous He-gas flow cryostat (Oxford
instruments) working in the temperature range from liquid helium
to room temperature.

\section{Experimental Results and Discussion}

\subsection{Magnetic susceptibility}

\begin{figure}
\resizebox{0.95\columnwidth}{!}{\includegraphics{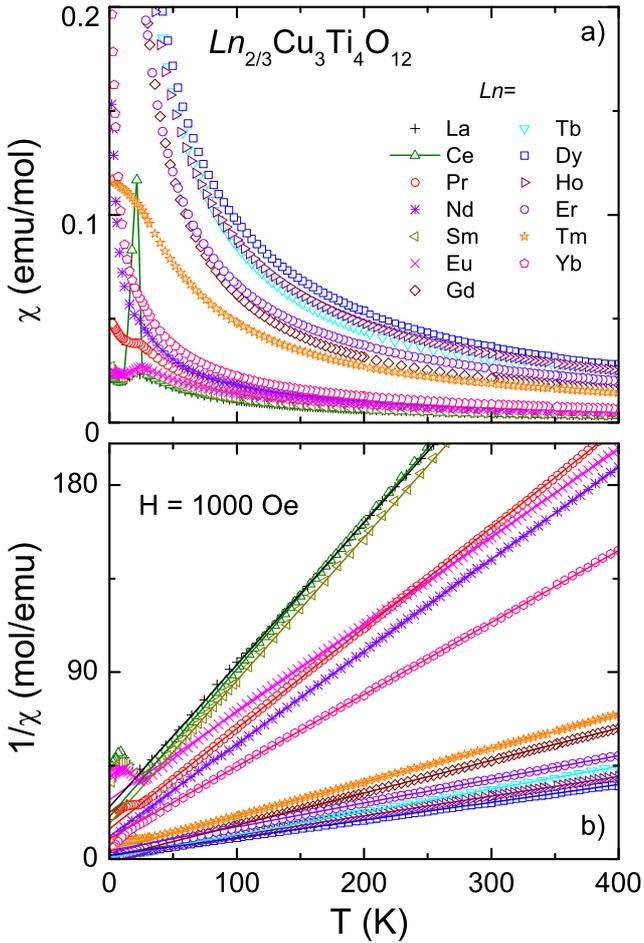}}
\caption{(Color online) Magnetic susceptibility measured at $H=1000$\,Oe of all \LnCTO compounds. a) Susceptibility vs.
temperature. b) Inverse susceptibility vs.
temperature. The solid lines in the lower frame correspond to
Curie-Weiss fits as described in the text. Symbols characterizing
the different $Ln$ compounds are indicated in the upper frame.}
\label{Fig4}
\end{figure}

Figure~\ref{Fig4} depicts the magnetic susceptibility as function
of temperature for all lanthanide compounds from La to Yb
in an external magnetic field
of 1000\,Oe and between 1.8\,K and 400\,K. The upper
frame shows the susceptibility, the lower frame the inverse susceptibility. The symbols
characterizing the different compounds are indicated in the upper
frame. Starting the discussion with the lower frame first, one observes
that all compounds exhibit a well defined Curie-Weiss
like susceptibility for $T > 50$\,K. Three groups
of lines with different slopes can be distinguished: The samples with $Ln =$ La, Ce, Sm
exhibit the steepest slopes. In these cases the transitions into
long-range AFM order are clearly visible close to 25\,K.
Naturally, for $Ln =$ La the slope is solely determined by the
spin $S = 1/2$ of the copper ions. The approximate
coincidence of the susceptibilities of the La and the Ce compounds
demonstrates that the majority of cerium ions is in the
non magnetic $4f^0$ state. For Sm$^{3+}$ the effective magnetic moment
is small compared to that of Cu$^{2+}$, therefore, the three copper ions
per formula unit dominate the susceptibility of this oxide.
The lowest slopes are found for the
heavy rare earths ($Ln =$ Tm, Gd, Er, Tb, Ho and Dy). For these
compounds the large moment of the rare-earth ion governs the susceptibility down to
the lowest temperature and even the AFM transition of the copper
ions is hidden by the large paramagnetic susceptibility.
Finally in between these two extremes, we find the compounds ($Ln =$ Pr, Nd, Eu, and Yb) where
the rare-earth moment (only $2/3$ per formula unit) and the copper
spins (3 Cu ions per unit cell) are of comparable magnitude.


Turning now to the upper frame we find two further significant
differences in the temperature dependent magnetic susceptibilities
of $Ln_{2/3}$Cu$_3$Ti$_4$O$_{12}$. The Tm compound reveals a saturation
towards low temperatures indicating the existence of a
non-magnetic ground state. This situation sometimes occurs for Tm
and Pr compounds in a cubic crystal field. In these cases the
crystal-field splitting of the rare-earth moment leads to a non-magnetic singlet ground state. Thus, the susceptibility at
low temperatures is of Van-Vleck type due to the
polarization of the singlet ground state \cite{Cooper1967}.
Secondly, \CeCTO exhibits a much more significant and enhanced
anomaly at the AFM ordering transition of the copper spins.

\begin{figure}
\resizebox{0.95\columnwidth}{!}{\includegraphics{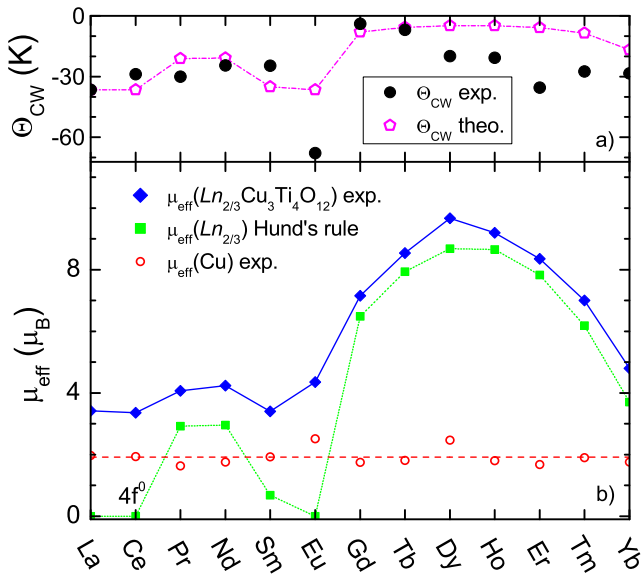}}
\caption{a) Experimental Curie-Weiss
temperatures (full black circles) of \LnCTO as determined from
high-temperature Curie-Weiss fits as shown in Fig.~\ref{Fig4} compared to calculated
Curie-Weiss temperatures (empty red circles) as described in the
text. The dash-dotted line is drawn to guide the eye. b) Effective moments as determined from the paramagnetic
high-temperature behavior (full blue diamonds). Calculated moments
of the rare-earth ions according to Hund's rule coupling (full
green squares) are presented in addition to the calculated
effective Cu moments for the \LnCTO compounds (empty red symbols).
Solid and dotted lines are drawn to guide the eye.
The dashed line is the average effective copper moment.}
\label{Fig5}
\end{figure}

For a more quantitative analysis we fitted the susceptibility
data of all compounds for $50 < T < 400$\,K by a Curie-Weiss law
\begin{equation}
\chi = \frac{N_{\rm A} \mu_{\rm eff}^2}{3 k_{\rm B}(T-\Theta_{\rm CW})} + \chi_0
\end{equation}
(with Avogadro number $N_{\rm A}$ and Boltzmann constant $k_{\rm B}$). In these fits we
included also a temperature independent term $\chi_0$ representing possible
diamagnetic behavior or Van Vleck type paramagnetism. The fits yielded positive Van Vleck like contributions of the order
of 10$^{-4}$ - $10^{-3}$\,emu/mol for all samples. Compared to the over-all size of
the susceptibility as shown in the upper frame of Fig.~\ref{Fig4},
these contributions seem to be rather negligible. The resulting
Curie-Weiss temperatures, $\Theta_{\rm CW}$, and paramagnetic effective moments,
$\mu_{\rm eff}$, are depicted in Figure~\ref{Fig5}. The
theoretically expected effective paramagnetic moments for \LnCTO are calculated via
\begin{equation}
\mu_{\rm eff}^2 = \frac{2}{3} \mu_{\rm eff}^2(Ln) + 3\mu_{\rm eff}^2({\rm Cu}).
\end{equation}
In the lower frame of Fig.~\ref{Fig5}, the measured values (full blue diamonds) were
analyzed using the following procedure: we calculated the
rare-earth moments according to Hund's rule coupling assuming that
at high temperatures all crystal-field levels are occupied and the
rare-earth ions exhibit their full moment. These calculated values
are indicated as full (green) squares. In these calculations we
assumed a non-magnetic $4f^0$ state for the cerium compound. These
values were subtracted from the measured effective moments to
give the average spin moment per copper ion over the complete
series of lanthanide compounds (full red circles). As can be seen,
these values slightly scatter around roughly 2.0~$\mu_{\rm B}$ yielding an
average effective moment of 1.920~$\mu_{\rm B}$ per copper ion.
This results in a $g$ value of 2.22 which is enhanced by
spin-orbit coupling with respect to the spin--only value as typical
for Cu--$3d^9$ systems. From this figure it is evident that the
assumption of non-magnetic cerium describes the results
satisfactorily.


The experimental Curie-Weiss temperatures depicted in the upper frame of Fig.~\ref{Fig5} have to be compared to the theoretically expected values, which for a two-sublattice antiferromagnet can easily be calculated using
\begin{equation}
\Theta_{\rm CW}  = \frac{2 \lambda_{AB} C_A C_B - \lambda_{AA} C_A^2 -
\lambda_{BB} C_B^2}{C_A + C_B}.
\end{equation}
Here $C_A$ and $C_B$ are the Curie constants of the $A$ (Cu) and of the
$B$ (rare-earth ions) sublattice, respectively. $\lambda_{ij}$ are the effective
coupling constants within the sublattices $A$ and $B$ and in
between these two sublattices. As first approximation it is
reasonable to assume that in \LnCTO the rare-earth ions are fully decoupled
yielding $\lambda_{BB} = 0$ and that also the coupling between the Cu
and lanthanide ions is weak, resulting in $\lambda_{AB} \sim 0$.
In this case the Curie-Weiss temperature is only given by
$\Theta_{\rm CW} = - \lambda_{AA} C_A^2 /(C_A + C_B)$.
In the La compound with only one sublattice one obtains
$\Theta_{\rm CW} = - \lambda_{AA} C_A = -37.0$\,K, very close to the
values found in \CCTO \cite{Krohns2009}. Calculating the
Curie-Weiss temperatures for the complete lanthanide series using
the effective moments for the rare-earth ions as indicated
in the lower frame of Fig.~\ref{Fig5}, gives the magenta open pentagons which
are plotted in the upper frame of Fig.~\ref{Fig5}. The agreement
is satisfactory with the exception of the Eu compound and the
heavy rare earths. It seems that in the latter case the coupling
between the rare-earth and the copper spin system cannot be fully
neglected.

For \EuCTO we additionally found an enhanced Curie-Weiss temperature $\Theta_{\rm CW} = -68$\,K and a significantly smaller slope than for La$_{2/3}$Cu$_3$Ti$_4$O$_{12}$,
although the magnetic moment of Eu$^{3+}$ vanishes in the ground state with
zero total moment $J=0$. This is a general observation for magnetic compounds
containing Eu$^{3+}$, like e.g. EuMnO$_3$ \cite{Hemberger2007}, and results from the thermal occupation
of the excited $J=1$ state of the europium ion yielding a temperature-dependent
van-Vleck contribution of the susceptibility at elevated temperatures, which effectively enhances
paramagnetic moment and Curie-Weiss temperature.

\subsection{Specific heat}

\begin{figure}
\resizebox{0.95\columnwidth}{!}{\includegraphics{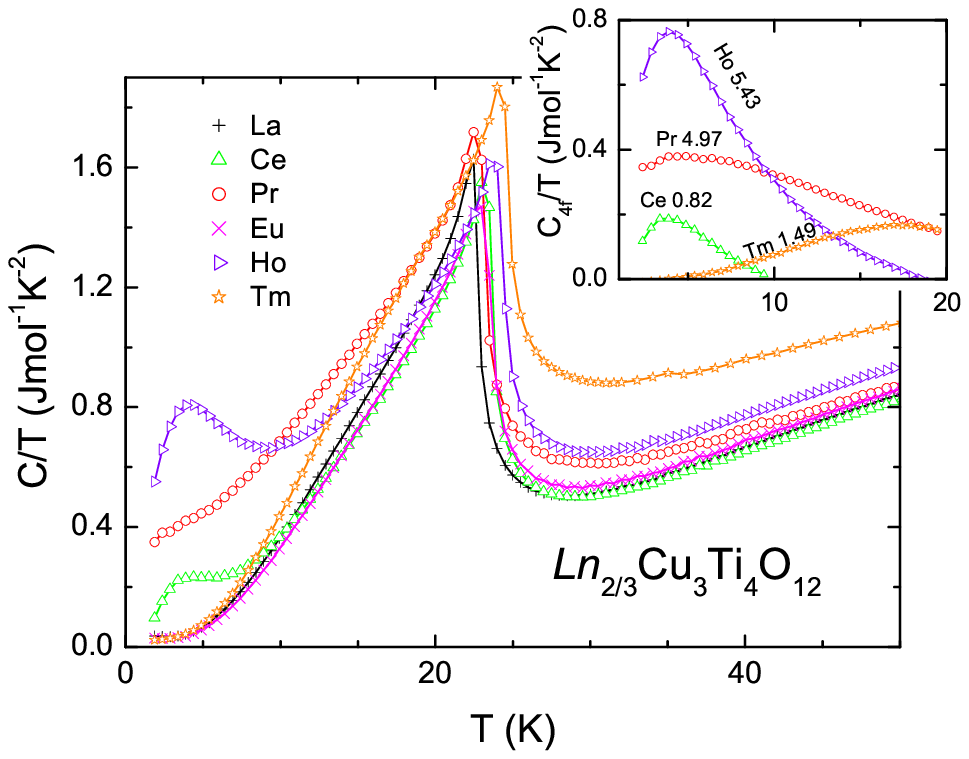}}
\caption{Heat capacity of \LnCTO with $Ln =$ La, Ce, Pr, Eu, Ho, and Tm the heat capacity is plotted as $C/T$ \textit{vs}
temperature. Symbols are indicated within the frame. The solid
lines are drawn to guide the eye. Inset: Rare-earth derived magnetic heat
capacities of \LnCTO for $Ln =$ Ce, Pr, Ho, and Tm, plotted as
$C/T$ \textit{vs} $T$. To determine the contribution of rare-earth ions the heat
capacity of \LaCTO has been subtracted from the measured heat
capacities shown in the main frame. The entropies due to
the crystal-field excitations have been estimated for the
temperature regime $2 < T < 20$\,K and are indicated in the figure
in units of J\,mol$^{-1}$\,K$^{-1}$.}
\label{Fig6}    
\end{figure}

For a number of \LnCTO compounds we measured the heat capacity
between 2\,K and 50\,K. Besides La$_{2/3}$Cu$_3$Ti$_4$O$_{12}$, we selected representative light
and\linebreak[4] heavy rare-earth compounds which behave rather unusual in the
temperature dependent magnetic susceptibility. In addition we
wanted to check the magnetic ordering temperatures, as these can
hardly be identified in the compounds with heavy rare earths (see
Fig.~\ref{Fig4}). Figure~\ref{Fig6} shows the measured heat
capacities for \LnCTO with $Ln =$ La, Ce, Pr, Eu, Ho, and Tm plotted
as $C/T$ vs. $T$. At first sight it is evident that all
compounds reveal antiferromagnetic order close to 25\,K similar to
\CCTO \cite{Kim2002}. At this temperature the copper
spins with $S = 1/2$ undergo long-range AFM order. It is
interesting to note that the ordering temperatures shift from 22.5\,K for the La to 25\,K for the Tm compound and it seems
that the heavy rare-earth compounds exhibit slightly higher
ordering temperatures. Obviously the heavy rare-earth compounds
exhibit non-vanishing exchange between the copper spins and the
$4f$ moments, a fact that also seems to be supported by the
evolution of the Curie-Weiss temperatures documented in the upper
frame of Fig.~\ref{Fig5}. An alternative explanation could be found in the lanthanide contraction slightly enhancing the superexchange between neighboring copper ions. 

In some of the systems crystal-field excitations can easily be
recognized. \HoCTO and astonishingly \CeCTO exhibit a well defined
hump at 5\,K indicating low-lying crystal-field excitations. In
case of the cerium compounds this has been rather unexpected as
from the diffraction and susceptibility experiments the cerium ion
has been characterized as $4f^0$, i.e. non magnetic. However,
Fig.~\ref{Fig6} proves that a certain fraction of cerium ions is definitely
in the $4f^1$ state. The anomaly for $Ln=$ Ho is significantly larger than for the Ce. Therefore we conclude that only a small fraction is in the Ce$^{3+}$ state. The heat-capacity
contributions due to the crystal-field excitations of \PrCTO and of
the singlet-ground state in \TmCTO extend over larger temperature
ranges, i.e. from 2\,K to almost 40\,K and from 10\,K to
temperatures well above 50\,K, respectively. Of course, no
$4f$-derived crystal-field excitations are visible in the
compounds with non-magnetic $A$-site \LaCTO and Eu$_{2/3}$Cu$_3$Ti$_4$O$_{12}$.


To get a more quantitative estimate of the $4f$ derived
contributions to the heat capacity for all compounds with
rare-earth crystal-field excitations, we subtracted the specific heat
of La$_{2/3}$Cu$_3$Ti$_4$O$_{12}$. This accounts for the pure phonon density of states and for
the heat capacity of the ordering copper spin system. The results
are documented in the inset of Fig.~\ref{Fig6} and represent an estimate of the heat capacity due to the $4f$ moments. Here we
show $C/T$ vs $T$ between 2\,K and 20\,K for \LnCTO with $Ln =$ Ce,
Pr, Ho, and Tm. The corresponding entropy values are indicated
at the plotted anomalies in units of J\,mol$^{-1}$\,K$^{-1}$.
Having in mind that in the regular structure only $2/3$ of the $A$
sites are occupied by rare-earth ions, the entropies correspond to
triplet states for the Ho and Pr compounds and roughly to a
singlet-singlet transition for the Tm compound. Assuming that the
crystal-field anomaly for \CeCTO corresponds to a transition within
a doublet, we conclude that about 10\% of the $A$
sites would be occupied by trivalent cerium, 40\% by tetravalent
cerium and 50\% would remain empty. Note that to obtain charge neutrality,
in this case the Ce occupation of the $A$ site has to be about 55\%.
Of course we cannot exclude that these crystal-field contributions result
from another cerium-containing phase. However, a value of 10\% impurity phase
seems to be too large, specifically when inspecting the
diffractogram as indicated in Fig.~\ref{Fig1}.

\subsection{Electron spin resonance}


\begin{figure}
\resizebox{0.95\columnwidth}{!}{\includegraphics{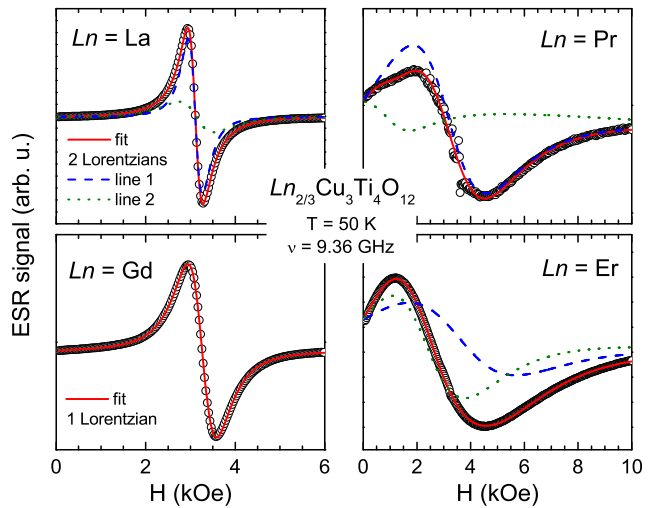}}
\caption{(Color online) Selected ESR-Spectra of \LnCTO taken at 50~K with $Ln =$ La, Gd, Pr, and Er. Solid lines indicate the fits, dashed and dotted lines illustrate the different contributions for fits with two Lorentzians.}
\label{spectra}    
\end{figure}

A complementary and microscopic access to the magnetic properties is provided via
electron spin resonance (ESR) which detects the microwave
absorption due to magnetic dipolar transitions between the Zeeman
levels of the magnetic ions. The ESR spectra are taken at constant
microwave frequency $\nu \approx 9.4$\,GHz dependent on the external static
magnetic field $H$. Due to the lock-in technique with field
modulation the field derivative of the absorption signal is
recorded. Figure~\ref{spectra} shows representative examples of ESR spectra
obtained in \LnCTO with different rare-earth ions: Four characteristic groups can be distinguished:
(a) for $Ln = $ La, Eu, Ce$^{4+}$, and Tm the signal is solely due to Cu$^{2+}$ spins as the rare-earth spin vanishes in the ground state; (b) in the case of $Ln = $ Gd the gadolinium spin dominates the absorption; (c) for $Ln = $ Pr and Nd the Cu$^{2+}$ signal dominates the spectrum, while the $Ln$ signal is weak; (d) both ions yield sizable contributions to the absorption in \ErCTO and the remaining compounds.

\begin{figure}
\resizebox{0.95\columnwidth}{!}{\includegraphics{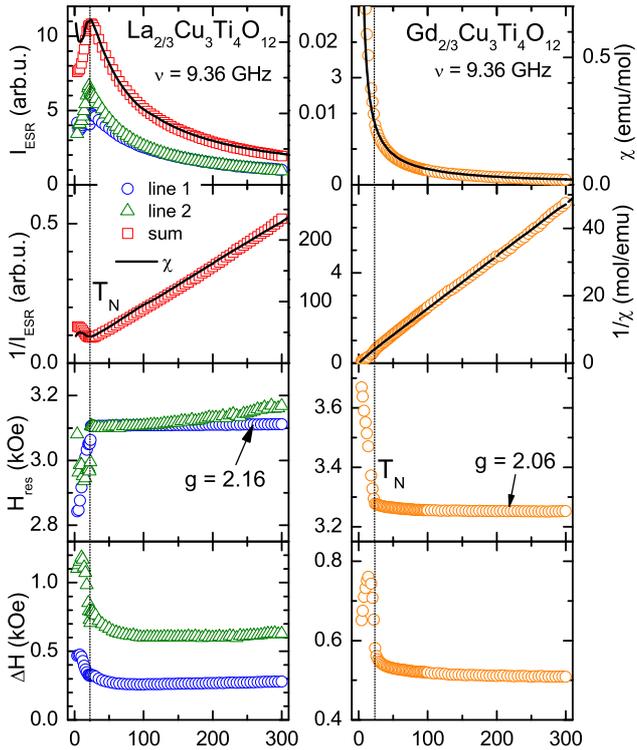}}
\caption{(Color online) Temperature dependence of ESR intensity (symbols) compared to SQUID susceptibility (solid lines), inverse intensity and susceptibility, resonance field and linewidth in \LaCTO (left frame) and \GdCTO (right frame).}
\label{LaGdESR}    
\end{figure}

To start with group (a) the spectra of \LaCTO (with La$^{3+}$ in
a $4f^0$ electron configuration) consist of a symmetric resonance line in the
paramagnetic regime, but reveal a clear asymmetry in the
magnetically ordered phase. The signal is well described by
two Lorentz curves of different linewidth but comparable intensity
and approximately the same resonance field corresponding to a
$g$ value $g=h\nu/\mu_{\rm B}H_{\rm res} \approx 2.16$, typical for
Cu$^{2+}$ where the orbital momentum is nearly quenched by the
ligand field \cite{Abragam1970}. The temperature dependence of
the fit parameters -- intensity, resonance field, and linewidth --
is depicted on the left side of Fig.~\ref{LaGdESR}. In the paramagnetic regime the
ESR intensities of both contributions follow very similar
Curie-Weiss laws, which sum up to the static susceptibility
probed by SQUID measurements. The antiferromagnetic transition
at $T_{\rm N}$ appears as a sharp kink followed by a drop of the
intensity to lower temperatures again in agreement with the static
susceptibility. The resonance fields $H_{\rm res}$ remain approximately constant
in the paramagnetic regime, but clearly shift to lower fields below
$T_{\rm N}$ due to internal fields in the ordered phase. At high
temperatures the linewidth $\Delta H$ attains values of about 250~Oe
and 650~Oe for the narrow and the wide contribution, respectively.
On approaching magnetic ordering the linewidth increases only
slightly driven by critical spin fluctuations but changes more pronounced
having passed $T_{\rm N}$ into the ordered phase.


The linewidth behaves similar to that observed in\linebreak[4]CaCu$_3$Ti$_4$O$_{12}$,
which was recently investigated in detail by Pires {\it et al.}, who performed
systematic annealing experiments using O$_2$ or Ar atmosphere \cite{Pires2006}.
In that work it was found that the main source of line broadening stems from
the dipolar interaction between neighboring Cu$^{2+}$ spins as well as with
some Ti$^{3+}$ spins generated by oxygen vacancies, whereby the exchange
interaction results in strong exchange narrowing of the signal. As grown or O$_2$ annealed
CaCu$_3$Ti$_4$O$_{12}$ samples exhibit a linewidth of about 40\,Oe, which results from Cu--Cu
dipolar broadening, only, where practically no oxygen vacancies exist. But Ar
annealing increases the linewidth by a factor of 5, because oxygen is extracted
from the sample leaving paramagnetic Ti$^{3+}$ instead of diamagnetic Ti$^{4+}$
switching on the dipolar interaction between Cu$^{2+}$ and Ti$^{3+}$, whereby the
Cu--Ti distance is significantly smaller than the Cu--Cu distance.
In \LnCTO probably disorder plays an additional role for the line broadening,
because only $2/3$ of the lanthanide sites are randomly occupied leading to different local coordination of the Cu spins. This may be the main reason for the necessity to fit the ESR signal by the superposition of two Lorentz lines of comparable intensity but different linewidth, which approximately accounts for the spatial fluctuations of the local interactions due to disorder. Because of the all over cubic crystal symmetry and the dominant exchange coupling between Cu sites of different local anisotropy axis, the usual explanation for the existence of two signal components in the powder pattern at $g$ values $g_{||}$ and $g_{\perp}$ for the magnetic field applied parallel and perpendicular to the local anisotropy axis can be ruled out.

A similar behavior like in the La compound is obtained for the Eu system,
where the rare-earth spin is zero in the ground state due to Hund's rule. Again the spectra are best described by two lines with slightly lower $g$ values 2.12 and 2.14 and comparable linewidths of 490~Oe and 260~Oe, respectively. In the Tm compound, which exhibits a zero-spin ground state at low temperatures due to the crystal-field splitting, even one Lorentz line nicely fits the ESR spectrum with $g \approx 2.1$ and $\Delta H \approx 750$~Oe in the paramagnetic regime. In \CeCTO the signal is satisfactorily described by a broad single resonance line at $g \approx 2.24$, which increases in width from 2 kOe at room temperature to 3 kOe just above $T_{\rm N}$. In the latter two compounds the signal is additionally broadened by the dipolar interaction with the Tm$^{3+}$ spins of the excited states and with about 10\% Ce$^{3+}$ spins detected also by specific heat, respectively.


For Gd$_{2/3}$Cu$_3$Ti$_4$O$_{12}$ -- case (b) -- the ESR spectrum is well described by a single Lorentz curve in the whole temperature range. The resonance field at $g = 2.06$ and linewidth behave similar to La$_{2/3}$Cu$_3$Ti$_4$O$_{12}$, with a distinct kink at $T_{\rm N}$, however the ESR intensity does not show any anomaly at $T_{\rm N}$. This signal has to be attributed mainly to the Gd$^{3+}$ spins $S=7/2$, which stay paramagnetic even in the antiferromagnetically ordered regime and, hence, follow the Curie-Weiss law down to low temperature, while resonance field and linewidth are sensitive parameters to the magnetic ordering of the Cu system. The Cu signal cannot be separated from the Gd signal, because its $g$ value nearly coincides with Gd and the linewidth is comparable, but its intensity is about a factor of 5 weaker than that of the Gd signal. Nevertheless, the $g$ factor of 2.06 indicates the admixture of the copper signal, because in insulating matrix Gd$^{3+}$ is expected to reveal a $g$ factor 1.993 \cite{Abragam1970}, whereas $g = 2.16$ is observed, if only Cu$^{2+}$ contributed to the resonance line.


\begin{figure}
\resizebox{0.95\columnwidth}{!}{\includegraphics{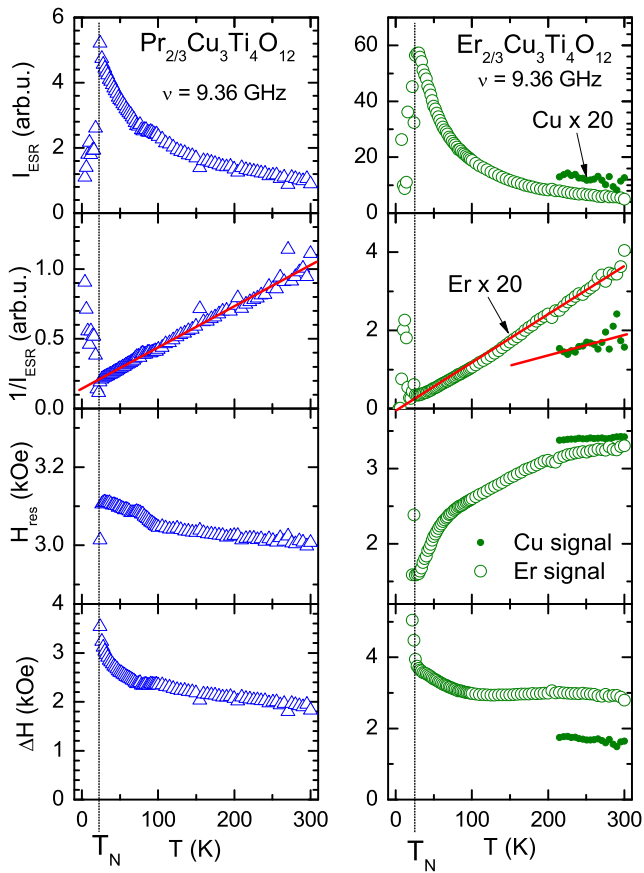}}
\caption{(Color online) Temperature dependence of ESR intensity, inverse intensity, resonance field and linewidth in \PrCTO (left frame) and \ErCTO (right frame). Solid lines in the inverse-intensity plots indicate the respective Curie-Weiss laws. For the Er compound the (inverse) intensity data of the (erbium) copper line have been magnified by a factor of 20.}
\label{PrErESR}    
\end{figure}

Focusing on group (c) represented by \PrCTO in the left column of Fig.~\ref{PrErESR}, the ESR spectra are satisfactorily described by a single Lorentz line at resonance fields corresponding to $g \approx 2.2$ with the linewidth increasing from about 2~kOe at room temperature up to 3~kOe and 4~kOe at $T_{\rm N}$ for \PrCTO and Nd$_{2/3}$Cu$_3$Ti$_4$O$_{12}$, respectively, plus a weak signal at a resonance field of about 0.7 kOe. Regarding the $g$-value the strong line can be ascribed to Cu$^{2+}$. Moreover, its inverse intensity yields a Curie-Weiss temperature of about $\Theta_{\rm ESR} \approx -40$~K, clearly below the SQUID value, which is influenced by the Curie contribution of the rare-earth spins. The intensity of the low-field line is by far too weak to account for the full rare-earth spin and is approximately independent on temperature. Probably this signal results from a small amount ($\ll 1\%$) of ferromagnetic impurities and, therefore, will not be further discussed. The rare-earth spin obviously does not give any measurable ESR signal. Likely due to spin-orbit coupling the relaxation of the rare-earth spin is so fast that the signal is too broad to be observable.

Concerning group (d) a reasonable separation of the different contributions to the ESR signal is possible in \DyCTO and Er$_{2/3}$Cu$_3$Ti$_4$O$_{12}$, which is illustrated in the right column of Fig.~\ref{PrErESR}. The copper signal is well resolved at $g \approx 2$ above 200\,K. To lower temperatures the Er line dominates and itself exhibits a substructure due to the crystal-electric field which can be best fitted in terms of two Lorentzians, as visible in Fig.~\ref{spectra}. But also a single Lorentz line yields a satisfactory description of the signal and was used for the present evaluation, because for a detailed investigation of the resonance field and substructure of the Er signal single crystals would be necessary. Here the most important information obtained is the signal intensity. At room temperature the copper intensity is one order of magnitude smaller than that of the erbium signal, as expected from ratio of the effective paramagnetic moments between Cu$^{2+}$ and Er$^{3+}$ (cf. Fig~\ref{Fig5}). Regarding the inverse intensities it is clearly visible that the copper signal follows a Curie-Weiss law with sizable negative Curie-Weiss temperature $\Theta_{\rm ESR} \approx -75$\,K, while the erbium signal exhibits a pure Curie law. This again indicates that copper and rare-earth lattices can be independently treated. For the Ho and Sm compound the copper line can be extracted in a similar way to determine $\Theta_{\rm ESR}$, but the rare-earth contributions turn out to be more complicated probably due to the crystal-field splitting. Finally, for Tb and Yb one observes very broad spectra with a width $\Delta H$ of about 4 and 5~kOe, respectively, at resonance fields less than 2~kOe, which are superimposed by narrower lines of large amplitude at $g \approx 1.9$ due to small amounts of Ti$^{3+}$ ions. The copper and rare-earth contributions can be identified at high temperatures, but a reliable termination of $\Theta_{\rm ESR}$ is not possible.


\begin{figure}
\resizebox{0.95\columnwidth}{!}{\includegraphics{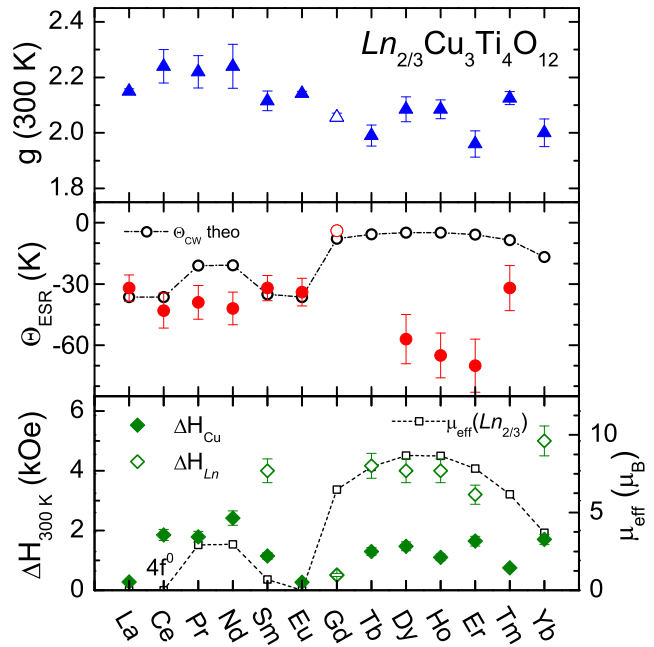}}
\caption{(Color online) Evolution of $g$-value, Curie-Weiss temperature $\Theta_{\rm ESR}$, and room-temperature ESR linewidth $\Delta H_{\rm 300 K}$ determined from the copper signal (solid symbols) in \LnCTO for different lanthanide ions. The open symbols indicate contributions of the rare-earth signal - for \GdCTO the copper signal could not be distinguished from the Gd-signal.}
\label{ESRparameter}    
\end{figure}

Figure~\ref{ESRparameter} summarizes the relevant results of the ESR measurements in \LnCTO dependent on the lanthanide ion. In all compounds with exception of\linebreak[4]\GdCTO the copper signal could be identified to determine the $g$-value and Curie-Weiss temperature $\Theta_{\rm ESR}$ of the copper sublattice. The $g$-values vary around an average value $g_{\rm Cu} = 2.18$ in reasonable agreement with the susceptibility data, whereby it is slightly enhanced for the Ce, Pr, and Nd system but smaller for the heavy rare-earth compounds. The Curie-Weiss temperatures $\Theta_{\rm ESR}$ are in general lower than those determined from the SQUID measurements, because they are not reduced by the Curie contribution of the rare-earth spins. Only for \GdCTO the $g$-value and Curie-Weiss temperature is dominated by the gadolinium spins, because their $g$-value roughly coincides with the copper spins. Note that especially for the Eu compound we find the proper Curie-Weiss temperature  $\Theta_{\rm ESR}=-34$\,K expected for the case of zero-spin Eu$^{3+}$, because the van-Vleck contributions of europium are not detected by ESR. Moreover, we observe larger absolute values of the Curie-Weiss temperature as compared to \LaCTO in the heavy rare-earth compounds indicating some influence of the rare-earth ions on the exchange interaction in the copper system, although the ESR signal of the rare-earth spins can be identified independently, which proofs the general independence of the magnetic sublattices. Finally, the copper linewidth taken at room temperature shows a qualitative similarity with the effective paramagnetic moment, i.e. the local dipolar fields of the rare-earth spins enhance the spin-spin relaxation. For the heavy rare-earth compounds the enhancement seems to be smaller, probably due to the fact that the exchange-narrowing is larger than for the light rare-earth systems as anticipated from the larger absolute values of the Curie-Weiss temperature. The rare-earth signal can be mainly identified in the heavy rare-earth systems, but due to its huge linewidth of about 4~kOe and underlying crystal-field splittings for its deeper analysis high-field and -frequency ESR as well as single crystals would be necessary.

\section{Conclusion}
In this work we presented a detailed structural, magnetic, and
thermodynamic characterization of \LnCTO where $Ln$ stands for
the lanthanides including La, Ce, Pr, Nd, Sm, Eu, Gd, Tb, Dy, Ho,
Er, Tm, and Yb. All polycrystalline samples crystallize in space
group $Im\bar{3}$. Most of the compounds were x-ray single phase. Traces of impurity phases ($< 2$\%) were detected in the Gd, Er,
and Tm compounds. Larger volume fractions of impurities ($\sim
5$\%) were identified for $Ln=$ Ce and Yb. The lattice
constants of the cubic cell nicely follow the lanthanide
contraction, with the exception of Ce$_{1/2}$Cu$_3$Ti$_4$O$_{12}$. The significantly too
small lattice constant indicates the presence of the smaller Ce$^{4+}$ ion, resulting in a 50\% occupation of the $A$ site only, to achieve charge neutrality.

The magnetic susceptibilities can nicely be described with a
two-sublattice model, with the copper spins $S = 1/2$ ordering at
25\,K and the rare-earth moments remaining paramagnetic down to
the lowest temperatures. The susceptibility of \TmCTO saturates at
low temperatures indicative for a non-magnetic ground state of the
crystal-field split $4f$ manifold. In all systems the rare-earth
ions are fully decoupled from each other. In the compounds with heavy rare-earth
ions a small coupling seems to exist between the copper spins and
the rare-earth ions. The copper ions are in a $3d^9$
configuration and their $g$ value is about 10\% enhanced with respect to 
the free electron value due to spin-orbit coupling.

Heat capacities between 2\,K and 50\,K were investigated for $Ln
=$ La, Ce, Pr, Eu, Ho, and Tm. All compounds reveal a clear anomaly
close to 25\,K indicative for antiferromagnetic ordering of the copper
spins. The ordering temperatures of the heavy rare-earth compounds
are slightly shifted towards higher temperatures ($\sim 2$\,K).\linebreak[4]
Crystal-electric field excitations have been detected for the Pr,
Ho, and Tm compounds. The former two can be interpreted due to a low-lying triplet. \TmCTO with singlet ground-state magnetism reveals a
singlet-singlet transition close to 20\,K. A small crystal-field
contribution was also identified in Ce$_{1/2}$Cu$_3$Ti$_4$O$_{12}$. Whether this contribution is
due to an impurity phase or Ce exist in both valences (Ce$^{3+}$ and Ce$^{4+}$) is still unclear in this compound.

Electron spin resonance allowed to identify the copper signal in most of 
the \LnCTO compounds and to extract the corresponding $g$-values and 
Curie-Weiss temperatures $\Theta_{\rm ESR}$ of the copper sublattice 
complementary to the SQUID measurements. Moreover, the onset of magnetic 
order is clearly observed in the ESR parameters in agreement with the 
specific-heat experiments. The copper linewidth is strongly affected by 
disorder due to the incomplete occupation of the $A$ place. Moreover, a 
qualitative relation with the effective rare-earth moment is observed, 
indicating the spin-spin relaxation channel between rare-earth spins 
and copper spins via the magnetic dipolar interaction. 

\begin{acknowledgement}
We thank Dana Vieweg for the SQUID measurements. This work was supported by the Commission of the European
Communities, STREP: NUOTO, NMP3-CT-2006-032644 and by the Deutsche Forschungsgemeinschaft (DFG) via
the Transregional Collaborative Research Center TRR 80 (Augsburg / Munich) and the Research Unit FOR 960. 
\end{acknowledgement}

%
%


\end{document}